# MILLIGRAM MASS METROLOGY FOR QUANTITATIVE DEPOSITION OF LIQUID SAMPLES

*Gordon A. Shaw*[*]

National Institute of Standards and Technology (NIST), Gaithersburg, MD, USA
* Corresponding author. E-mail address: gshaw@nist.gov

*Abstract* – Quantitative dispensing of liquids by mass becomes increasingly difficult as sample volume decreases. Whereas ml quantities of liquid can be directly weighed as dispensed on an analytical balance or using pycnometer methods, μl samples are more challenging due to their milligram-scale mass and rapid evaporation. Here, a method for quantitative determination of the mass of 1 μl droplets of liquid is proposed. The weighing of a microcapillary pycnometer combined with traceable calibration masses and corrections for sample evaporation allows determination of the droplet mass on the order of 1 mg with relative combined expanded uncertainty of 0.3 %.

## 1. INTRODUCTION

Mass metrology has long been used to quantify the amount of a liquid material. Although volume is frequently used for this purpose in laboratory settings, mass measurements are often the basis for the calibration of liquid volume [1-3]. If the quantity of fluid decreases towards the milligram level (approximately 1 μl in volume,) specific challenges arise. Although metal milligram calibration masses can be characterized with relative uncertainties in the range of $10^{-6}$, liquid masses in this range have typical uncertainties of 5 % [3] for an example application of gravimetric piston pipette volume calibration at 1 μl. Uncertainties over 1 % are acceptable for many industry applications, but requirements may differ for metrological systems. In apparatus designed to measure micro-flow, relative uncertainties in fluid mass measurement reach the level of $3 \times 10^{-5}$ [4], albeit, for 100 mg of sample. This higher level of precision may be necessary to quantify small quantities of dispensed material in law enforcement, domestic security, or fundamental science.

Gravimetric Droplet-on-Demand (GDoD) systems are particularly useful for depositing small quantities of liquid samples. These systems use a piezoacoustic resonator to eject a jet from a submillimeter aperture in a process akin to inkjet printing. By concurrently monitoring the reading of an ultra-microbalance, the total mass of dispensed liquid may be quantified [5]. Assuming the jetting process is well controlled, the same protocol used over the balance may be used to deposit the same amount of material onto a sample surface. The primary challenge in these experiments, as in many other liquid mass measurements, is accounting for evaporation. A 1 mg droplet of water will evaporate completely in the span of a few tens of seconds, depending on local conditions (approximately the same as the stabilization time for a high-precision balance.) The process of measuring the mass of an evaporating liquid is by necessity dynamic, in contrast with more typical static measurements associated with solid samples, which can cause further problems with obtaining stable balance readings. Another difference is that a liquid droplet cannot be weighed repeatedly the way a solid reference weight can. The droplet can only be dispensed once, and the mass must be determined at that point. This means the statistics of the measurement must be determined based on the statistics of the mass measurement procedure.

Over time, the analysis of GDoD deposition experiments has continued to evolve [6,7] with practitioners developing progressively more refined methods to account for the changing mass of their samples and developing models to estimate uncertainty in the mass dispensed for liquid volume from nl to μl. These studies share the approach where the mass of liquid is measured after it is dispensed. A different approach involves weighing a pycnometer before and after liquid is dispensed and has been used historically for applications such as radionuclide activity characterization [8]. The pycnometer approach has typically been applied with comparatively large amounts of sample, 10 milligrams or more. In the present work, we propose a microcapillary pycnometer measurement that allows quantification of milligram (μl) droplets by mass.

The buoyancy corrected mass for an object in air, $m_a$, is calculated using

$$m_a = m_t \left(1 - \frac{\rho_a}{\rho_X}\right), \quad (1)$$

where $m_t$ is the object's true mass, $\rho_X$ is its density, and $\rho_a$ is the density of air at the local temperature, pressure and humidity. A balance reading while weighing the object, $C$, is

$$C = \frac{m_a}{K}, \quad (2)$$

where $K$ is a proportionality constant determined by a calibration method, and is assumed to be linear within a specified uncertainty. For the dispensing of a μl drop, the balance reads out a measurement of the water mass in the capillary before, $C_0$, and after, $C_1$, the dispensing such that

$$C_0(t) = \frac{m_{t0}\left(1-\frac{\rho_a}{\rho_f}\right) + m_{tc}\left(1-\frac{\rho_a}{\rho_f}\right)}{K} + B_0(t), \quad (3)$$

$$C_1(t) = \frac{(m_{t0}-m_{td})\left(1-\frac{\rho_a}{\rho_f}\right)+m_{tc}\left(1-\frac{\rho_a}{\rho_f}\right)}{K} + B_1(t), \quad (4)$$

where $m_{t0}$ is the true mass of the fluid in the capillary as it is loaded from a large reservoir at time = 0, $m_{td}$ is the true mass of the droplet, $m_{tc}$ is the true mass of the capillary, $\rho_f$ is the density of the fluid, $\rho_c$ is the density of the capillary material and $B_0(t)$ and $B_1(t)$ are the measurement baselines which change slowly over time due to the combined effects of balance drift and evaporation of the fluid from the capillary (when present). A direct measurement of $C$ is typically not available as the droplet is being dispensed, since either the dispenser must be removed from the balance, or if the dispenser is mounted on the balance the mass will change too rapidly for the balance to follow. Here we can see if a linear fit is done to the time series of the balance reading, before and after the droplet is dispensed, the resulting intercept will be the first term in Eq. 3 and 4, and the second term will be the slope. If $C_d=C_0-C_1$, the differential measurement result is

$$C_d(t) = \frac{m_{td}\left(1-\frac{\rho_a}{\rho_f}\right)}{K} + B_0(t) - B_1(t). \quad (5)$$

In this fashion, the difference between the linear fits at a given value of $t$ yields the expected balance reading for the droplet of the mass at that time. It is also apparent that the buoyancy correction need only be applied to the measured mass of the dispensed droplet as long as the evaporation rate is small. A similar argument can be applied to the capillary dispenser, as its mass can be assumed to be approximately constant over the course of the experiment.

In order to determine $K$, a reference mass may be added after the droplet is dispensed. The balance reading after adding the reference is

$$C_2(t) = \frac{(m_{t0}-m_{td})\left(1-\frac{\rho_a}{\rho_f}\right)+m_{tc}\left(1-\frac{\rho_a}{\rho_c}\right)+m_{tr}\left(1-\frac{\rho_a}{\rho_r}\right)}{K} + B_2(t), \quad (6)$$

where $m_{tr}$ and $\rho_r$ are the true mass and density of the reference and $B_2(t)$ is the baseline. If $C_K = C_2 - C_1$,

$$C_K(t) = \frac{m_{tr}\left(1-\frac{\rho_a}{\rho_r}\right)}{K} + B_2(t) - B_1(t). \quad (7)$$

Combining Eq. 5 and 7 provides a measurement equation for measuring the droplet mass with the procedure,

$$m_{td} = \frac{m_{tr}b_r}{b_f}\frac{C_{dm}(t)}{C_{Km}(t)}, \quad (8)$$

Where $b_r=1-\rho_a/\rho_r$, $b_f=1-\rho_a/\rho_f$ $C_{dm}(t)=C_d(t)-B_0(t)+B_1(t)$ and $C_{Km}(t)=C_K(t)-B_2(t)+B_1(t)$, represent the predicted value of the differences between the linear fits to different areas of the balance timeseries data, as described below.

## 2. EXPERIMENT

The experimental apparatus consists of an ultra-microbalance (Mettler-Toledo XPR6U)[*] and a 5-µl capillary pipette. The balance digitally records mass measurements in evenly spaced time intervals of approximately one second, displaying the number of time steps elapsed. Prior to the experiment, the approximate mass of the empty capillary was measured by differential weighing using the internal calibration function of the balance while the balance was exercised for one half hour. Drummond Microcap microcapillary pipettes were then used to aspirate approximately 4 µl of 18 MΩ deionized water. Ultrapure water was chosen for initial experiments due to its well-characterized density. If Eq. 8 is to be used with another liquid, its density must be determined, and the uncertainty calculated as described below. Droplets were dispensed using a squeeze bulb adapted for use with the microcapillaries, which was removed during weighing. The measurement protocol is listed in Table 1.

Table 1. Experimental Protocol.

| Time Step (AU) | Procedural Step |
|---|---|
| 0 | Start data acquisition, fill capillary |
| 50 | Place full capillary on balance |
| 150 | Remove capillary and dispense droplet |
| 200 | Place capillary on balance |
| 300 | Remove capillary |
| 350 | Place capillary and 5 mg reference on balance |
| 450 | Remove capillary and 5 mg reference |
| 500 | Place capillary on balance |
| 600 | Remove capillary from balance |
| 650 | Place 5 mg reference on balance |
| 750 | Remove 5 mg reference |
| 800 | End |

## 3. RESULTS

Time series data are shown in Fig. 1 for one measurement trial. As weight is added or removed from the balance pan, larger transient forces can occur. As a result, large changes in

---

the measured mass can be seen at transition points in the data. These are normal for a manual weighing process. Linear fits are performed on the final 50 points of each portion of the data in which a fluid loaded capillary is present. It is found that approximately 30 s is required for the balance to regain equilibrium after loading or unloading, hence data within these times are excluded.

The terms $C_{dm}$ and $C_{Km}$ are determined by subtracting the linear fits in regions (1) and (2), and in regions (2) and (3) and (3) and (4) of Fig. 1, respectively. The actual time of the droplet dispensation is not known, other than that it is sometime between the unloading and replacement of the capillary. To accommodate this, $C$ values are determined over a range of times between regions (1) and (2), yielding a range of balance readings. This range of readings will be used as the temporal extrapolation uncertainty below.

Equation 8 provides a basis for a preliminary uncertainty calculation based on established models [9]. The measurement results and associated uncertainties are summarized in Table 2. All results are presented with coverage factor k = 2.

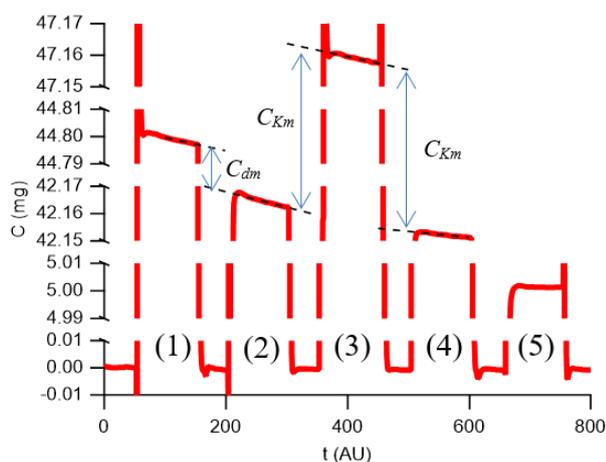

Figure 1. Experimental data from microcapillary pycnometer experiment time series (note broken y-axis), AU stands for arbitrary units. Red trace illustrates balance reading, $C$, dashed lines indicate linear fit to evaporation trend. From the measurement start, the filled pycnometer is placed on the balance at (1), is subsequently removed to dispense a drop and replaced at (2), removed again and replaced along with a reference mass at (3), removed again and finally replaced at (4). The reference mass is weighed at the end as a check standard (5). The extrapolation of the linear trend allows calculation of the expected change in the balance reading from the droplet ($C_d$) and the reference mass ($C_K$). $C_K$ is measured twice to ensure consistency.

Expanded uncertainty in $m_{tr}$ is evaluated based on the certified 3 µg tolerance of its E0 weight set class to be 2/3*tolerance = 2 µg. The air buoyancy correction terms were calculated using published data [11, 12].

Because no climate monitoring was carried out during measurement, the density of the air and water were calculated assuming a potential temperature range from 15 °C to 27 °C, relative humidity range from 0 % RH to 100 % RH, and barometric pressure range from 105 kPa to 87 kPa (which covers the highest to lowest barometric pressure readings ever recorded.) The laboratory density values of air and water were taken to be the mean of these two extrema, and the uncertainty is based on their difference. The laboratory temperature is typically 21±2 °C, and the humidity and barometric pressure range results in a very conservative uncertainty that covers all reasonable terrestrial conditions.

$C_{Km}$ and $C_{dm}$ are calculated by subtracting the linear fits from the time series weighing data as described above. Note that $C_{Km}$ is determined twice, and the mean value is reported in table 2. These values have two associated uncertainties. A type A (statistical) uncertainty is calculated from the linear least squares curve fit algorithm [12]. The slope and intercept value uncertainties ($k=2$) calculated from the data are used to define a minimum and maximum predicted value over the region between either (1) and (2) in Fig. 1 for $C_{dm}$ or between (2) and (3) and (3) and (4) for $C_{Km}$. This is done by adding or subtracting the uncertainties to the slope and intercept to produce a range of predicted values for the difference between the linear fits. The maximum in this range is taken as the extrapolation uncertainty in Table 2. For comparison, a separate analysis was performed using a t-test method used previously [6, 7]. This method produced a substantially lower uncertainty, however more data are required to validate this statistical methodology for the current experiment, so the more conservative uncertainty is used for the current work. The type B (non-statistical) temporal variation uncertainty results from uncertainty in the exact time of the droplet dispensing, and is described above.

A small correction for excess evaporation, $E$, is also added to $m_{td}$ to reflect extra evaporation that occurs during the dispensing process. To evaluate this correction, an experiment was performed wherein four water droplets were partially dispensed then re-aspirated into the capillary. Timeseries weighing of the capillary before and after the partial dispensing indicated a small decrease in the combined mass of the capillary and fluid. The mean and standard deviation of this decrease is used to calculate a value of 8.3±3.1 µg for $E$.

It was also apparent that mass transfer is possible as the pipette is transferred from the balance and inserted into the capillary squeeze bulb. It was found this could be minimized by repeatedly inserting and removing a separate capillary that was then discarded. Balance linearity is also expected to contribute negligibly to uncertainty over the small range of loads examined here.

Adding these uncertainty terms in quadrature yields a value and combined expanded uncertainty for $m_{td}$ of 2.6364(73) mg, yielding a relative combined expanded uncertainty of $2.8 \times 10^{-3}$. The experiment was repeated twice more and results are shown in Table 3.

## 3. DISCUSSION

The droplet mass indicated a volume of approximately 2.6 µl, and the relative uncertainty compares favorably with commercially available mass-based pipette calibration services in this volume range, even when considering the density variation due to temperature uncertainty. Buoyancy correction uncertainties could easily be reduced by an order of magnitude by monitoring laboratory conditions.

The reference masses used are off-the-shelf commercial E0 class masses. A careful calibration has the potential to reduce the uncertainty in their true mass values by two orders of magnitude [13].

The excess evaporation correction requires more advanced environmental control to reduce. An enclosure with a source of water vapor to elevate the local humidity would reduce the evaporation rate. In addition, the exact local conditions within the balance draft shield are not precisely known. The rate of evaporation is substantially less than previous work, however [7], helping to reduce any effects from local humidity changes. This also allows the balance feedback mechanism to more easily track the changes in mass since the sample evaporates more slowly from the 1 mm diameter capillary than the larger apertures used for the solvent reservoirs in other instruments. Nevertheless, an automated dispensing system still has distinct advantages in convenience relative to the hand pipetting of μl quantities of material. An automated system will also control the amount of deposited material more tightly. Although the mass dispensed from the capillary can be measured very precisely, the mass of the droplet varied from 0.9 mg to 2.6 mg in the three trials performed.

Table 2. Values and uncertainties for experimental terms ($k=2$). For unitless buoyancy compensation terms, $u$ is determined by multiplying relative term uncertainty by $m_{td}$.

| Term | value | $u$ (mg) | notes |
|---|---|---|---|
| $m_{tr}$ | 5.000 mg | 0.0020 | Reference mass |
| $b_r$ | 0.99986 | $4 \times 10^{-5}$ | Reference mass buoyancy compensation |
| $b_f$ | 0.99886 | $4 \times 10^{-4}$ | Fluid mass buoyancy compensation |
| $C_{Km}$ | 5.0021 mg | $2.9 \times 10^{-3}$ | Extrapolation uncertainty (type A) |
| | | $9 \times 10^{-4}$ | Temporal variation uncertainty (type B) |
| $C_{dm}$ | 2.6270 mg | $8 \times 10^{-4}$ | Extrapolation uncertainty (type A) |
| | 0.0005 mg | $4 \times 10^{-4}$ | Temporal variation uncertainty (type B) |
| $E$ | 0.0083 mg | $6.2 \times 10^{-3}$ | Excess evaporation |
| $m_{td}$ | 2.6364 mg | 0.0073 | Combined Expanded Uncertainty |

Nevertheless, the preliminary analysis reported above provides a means to measure the mass of a millimeter-sized droplet in a fashion traceable to the International System of Units (SI) with a defined uncertainty model. This provides a path forward for new type of small mass metrology that leverages the capability of precision chemical dilutions to provide a small quantity of a substance of interest within a solvent matrix of a much larger mass that can be readily quantified. The present technique may be particularly useful for experiments requiring the mass of samples deposited onto miniature sensors.

The accuracy of the technique has not been rigorously cross-checked; however, the internal calibration of the balance used in the experiment provides an ad-hoc process control monitor. The balance uses an automated system to calibrate its load cell with internal reference masses. With proper qualification, the balance reading itself can be considered a traceable measurement based on the internal calibration. This may be particularly useful when the use of extra reference weights is cumbersome. The value of $C_{Km}$ and the check weighing at the end of the experiment should provide an accurate value for the reference mass (within its tolerance). Table 3 shows the balance's internal calibration is consistent with the 5 mg reference mass, as was the case for all the droplet mass measurements. The check standard weighing also provided a sensitive process validation. The balance reading for the checkweight agreed with the value of the reference within its tolerance for each experiment except for the last one. In the last experiment, the capillary pipette was dropped when removing it from the balance pan. This disturbance in the experiment was sufficient to yield a balance reading outside of the reference mass's tolerance. This type of anomaly can be monitored as part of the weighing process, and the data from those samples excluded. Further work cross-checking this liquid sample weighing method against other traceable methods will be required to ensure the accuracy necessary to validate the technique fully as an SI-traceable measurement of liquid mass for metrological applications.

Table 3. Results and uncertainties ($k=2$) of droplet mass measurements.

| Trial | 1 | 2 | 3 |
|---|---|---|---|
| Value and $u_c$ | 2.6364(73) mg | 0.93750(79) mg | 1.1742(74) mg |
| Relative $u_c$ | 0.19 % | 0.61 % | 0.44 % |

## 4. CONCLUSIONS

A procedure for manual quantitative mass dispensing of milligram-scale droplets (of approximately μl volume) has been tested using deionized water as a model liquid sample. The preliminary measurement and uncertainty analysis suggest that droplet mass can be determined with relative combined expanded uncertainty of 0.3 %, and further improvement appears to be possible. Additional work will be required to validate the measurement approach; however, the use of an ultra-microbalance and calibrated weights provides a direct chain of traceability to the SI using readily available laboratory equipment.

## ACKNOWLEDGMENTS

Funding was provided by the NIST Innovations in Measurement Science (IMS) program under the TrueBq Project. Thanks to R. Michael Verkouteren, Patrick Abbott, Edward Mulhern and Zeina J. Kubarych for helpful discussion.